# Study on the effect of humidity and dust on leakage current of bulk micro-MEGAS detector[*]

WANG Bo(王波)[1;1)] ZHANG Yu-Lian(张余炼)[1, 3] QI Hui-Rong(祁辉荣)[1] LIU Jing(刘静)[1,2] ZHANG Xin-Shuai(张新帅)[1,2] ZHANG Tian-Chong(张天冲)[1] YI Fu-Ting(伊福廷)[1] OUYANG Qun(欧阳群)[1] CHEN Yuan-Bo(陈元柏)[1]

1) (*Institute of High Energy Physics, Chinese Academy of Sciences, Beijing 100049, China*)
2) (*University of Chinese Academy of Sciences, Beijing 100049, China*)
3) (*Lanzhou University, Lanzhou 730000, China*)

In this paper, the effect of humidity and dust trapped in avalanche region on leakage current of bulk micro-MEGAS detector is studied. Pyralux PC1025 layers of DuPont are introduced in bulk technique and micro-MEGAS detector with pillars of 300μm in diameter is fabricated. Leakage current is tested in air with different humidity. Silicon carbide powder and PMMA (polymethyl methacrylate) powder are added as dust to avalanche region. Leakage current with and without powder is tested in air and results are depicted in the same figure. Test results indicate that leakage current increases with both storage humidity and test humidity, and also increases when powder is introduced in avalanche region.

**Key words:** micro-MEGAS detector; leakage current; humidity; dust.

**PACS:** 29.40.Cs

## 1. Introduction

The micro-MEGAS detector was invented in the middle of 1990s [1]. micro-MEGAS detector consists of a drift electrode, a gap of a few millimeter thickness acting as conversion region (drift region), and a thin metallic mesh at typically 100μm distance from the readout electrode, creating the avalanche region (amplification region). The drift electrode and the amplification mesh are at negative high voltage potentials, and the readout electrode is at ground potential. The high voltage potentials are chosen such that the electric field in the drift region is a few 100V/cm and about 50kV/cm in the avalanche region. Charged particles traversing the drift space ionize the gas and electrons liberated by the ionization process drift towards the mesh. The mesh is transparent to most of the electrons as long as the electric field in the avalanche region is of the order of 100 times larger than the drift field. The electron avalanche takes place in the thin avalanche region, immediately above the readout electrode.

The bulk technique which was developed in Saclay-CERN around 2004 [2] is the most popular fabrication process of the avalanche structure. In this paper the pillars formed by PC1025 film are of 300μm in diameter and the interval is 2mm. Stainless steel woven mesh is used as avalanche electrode.

[*]Supported by National Natural Science Foundation of China (11275224)
1) corresponding author, E-mail:wangbo@ihep.ac.cn

Leakage current is an important parameter, which will affect the test when big enough, especially for large area detectors. Leakage current is discovered to be affected by both storage humidity and test humidity during our experiments. In this paper, leakage current is stored and tested in air with different humidity, and results are depicted in the same figure for comparison. Dust trapped in avalanche region also increases leakage current [3]. In this paper PMMA powder and silicon carbide powder are added as dust to avalanche region, and leakage current with and without powder is tested in air and results are depicted in the same figure.

## 2. Test and Result

The micro-MEGAS detector is kept in a shield to eliminate undesired signal during the tests. Leakage current is tested in air. Tests are carried with model 6517A electrometer/ high resistance meter of Keithley, and the voltage changes from 10V to 500V every 10V. The results of a detector are depicted in the same figure for comparison.

### 2.1 Test and result in different humidity

Five micro-MEGAS detectors are tested with different humidity, and one of them marked with 1# here is chosen to be discussed in this paper. Detector 1# is stored in the lab for two days with normal lab humidity of 40%RH ~ 60%RH, and then leakage current is tested with humidity of 58%, and the result is depicted line 1 in Fig.1. Leakage current is tested with humidity of 50%RH after stored in the lab for two days with lab humidity of 40%RH ~ 60%RH, and the result is depicted in line 2 in Fig.1. Then detector 1# is stored in drying tower for two days with humidity of 10%RH, and then tested with humidity of 50%RH, and the result is depicted in line 3 in Fig.1. At last detector 1# is tested with humidity of 10%RH in the shield after stored in drying tower for two days with humidity of 10%RH, and the result is depicted in line 4 in Fig.1. Storage humidity and test humidity of the four lines in Fig.1 is depicted in Table 1.The results in Fig.1 indicate that leakage current increases with storage humidity and test humidity. The other four detectors present similar performance during the test.

Table 1. Storage and test humidity of four lines in Fig.1.

|        | Storage humidity | Test humidity |
|--------|------------------|---------------|
| Line 1 | Lab humidity     | 58%RH         |
| Line 2 | Lab humidity     | 50%RH         |
| Line 3 | 10%RH            | 50%RH         |
| Line 4 | 10%RH            | 10%RH         |

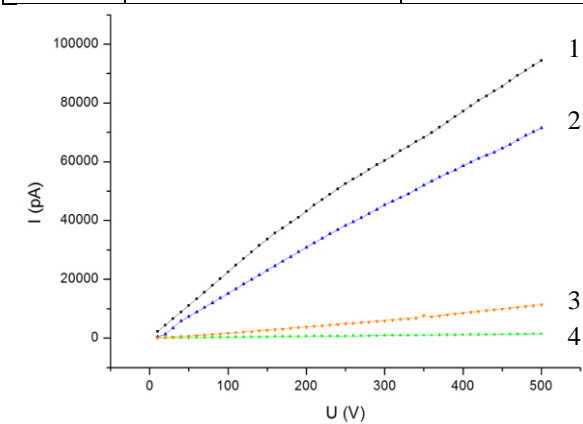

Fig.1. Leakage current with different humidity of detector 1#.

## 2.2 Test and result with powder

Silicon carbide powder whose average diameter is about 10μm is added as dust to the avalanche region of micro-MEGAS detector 2# and 3#, then the leakage current is tested in the shield with humidity of 10%, and the result is depicted in line 2 in both Fig.2 and Fig.3. Line 1 in both Fig.2 and Fig.3 is the result of leakage current without powder tested with humidity of 10%. As shown in Fig.2 and Fig.3, the introduction of silicon carbide powder increases leakage current of the micro-MEGAS detectors.

Another powder of PMMA whose average diameter is about 20μm is also introduced as dust to study the influence on the leakage current. The leakage current without powder of detector 4# and 5# is tested in the shield with humidity of 10%, and the result is shown in line 1 in both Fig.4 and Fig.5. Then PMMA powder is added to the avalanche region of detector 4# and 5#, and the leakage current is tested with humidity of 10%, and the result is depicted in line 2 in both Fig.4 and Fig.5. As shown in Fig.4 and Fig.5, the introduction of PMMA powder increases leakage current of the micro-MEGAS detectors.

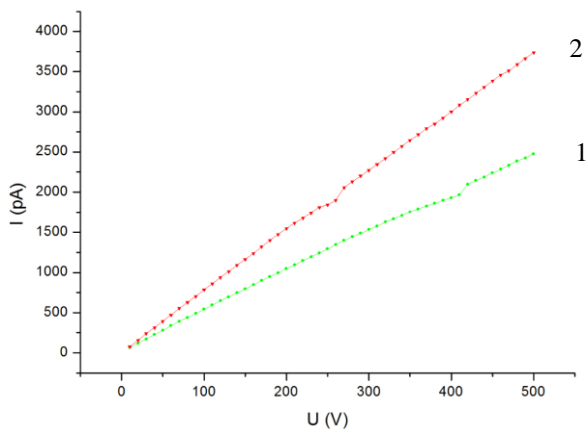

Fig.2. Leakage current with and without silicon carbide powder of detector 2#.

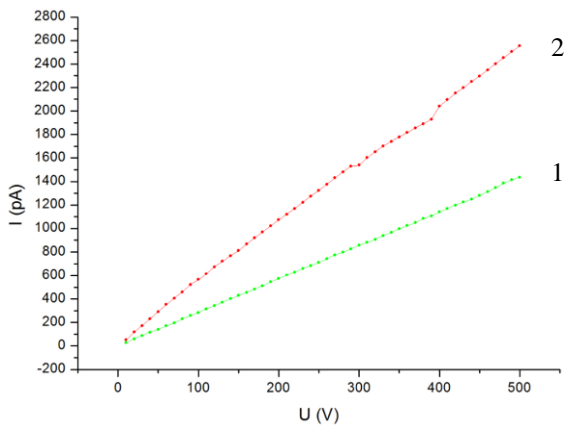

Fig.3. Leakage current with and without silicon carbide powder of detector 3#.

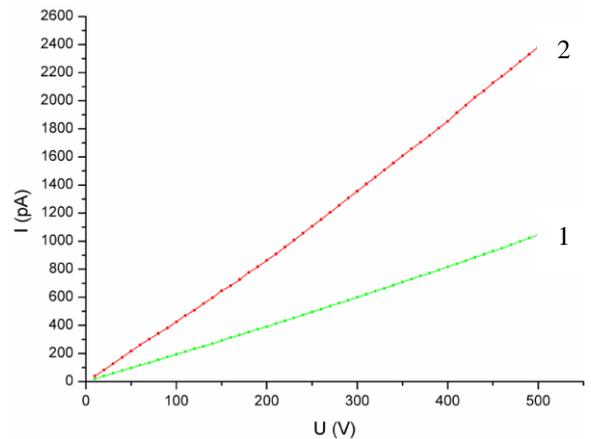

Fig.4. Leakage current with and without PMMA powder of detector 4#.

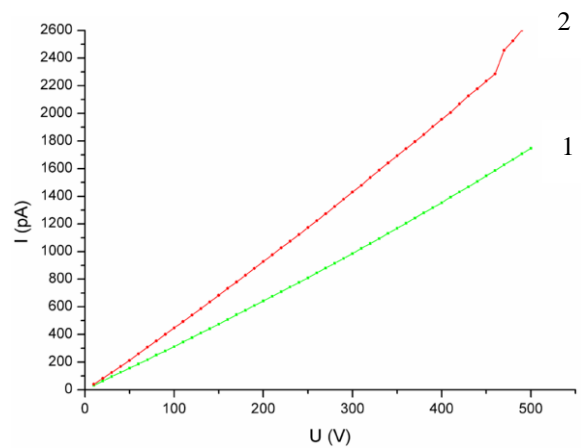

Fig.5. Leakage current with and without PMMA powder of detector 5#.

Both silicon carbide powder and PMMA powder are artificially added to the avalanche region. A little powder is scattered over the woven mesh, and then the detector is shaken and blown to make the powder drop into the avalanche region through the holes of woven mesh.

## 3. Discussion

The test results indicate that leakage current increases with test humidity comparing line 1 with line 2, and line 3 with line 4, as shown in Fig.1. It also indicates that leakage current increases with storage humidity comparing line 2 with line 3. We come to the conclusion from the test result that leakage current increases with both storage humidity and test humidity, and it is necessary to keep the environment as dry as possible during the storage and test process for a small leakage current.

It is indicated that the introduction of silicon carbide powder (Fig.2 and Fig.3) or PMMA powder (Fig.4 and Fig.5) in avalanche region increases the leakage current. However it is uncertain that which powder makes more contribution to the increase of the leakage current because of the uncertainty of the amount of powder, the different diameter of the powder, and the different dielectric constant, and so on. The contribution to the increase of the leakage current of each powder cannot be quantified because the powder is impossible to calculate and the distribution of the powder in avalanche region is unclear. Further experiments will be carried out to find out the relationship between the increase of leakage current and different powder, and the relationship between the increase of leakage current and diameter of each powder.

Luckily the size of the hole in woven mesh is about 40 μm, and dust powder larger than 40 μm cannot drop into the avalanche region, or it will lead to the worse according to ref[3].

## 4. Conclusion

Preliminary work indicates that leakage current increases with both storage humidity and test humidity, and also increases when powder is introduced in avalanche region, which give us the advice on the fabrication process and the test experiment. Experiments with different powder dust with different diameters will be carried out in the further.